\documentclass[letterpaper, 10 pt, conference]{ieeeconf}

\IEEEoverridecommandlockouts
\usepackage{cite}
\usepackage{amsmath,amssymb,amsfonts,amsthm}
\usepackage{upgreek}
\usepackage{algorithmic}
\usepackage{graphicx}
\usepackage{textcomp}
\usepackage{xcolor}
\usepackage{booktabs}
\usepackage{makecell}
\usepackage{csquotes}
\usepackage{hyperref}
\usepackage{siunitx}
\usepackage[nolist,nohyperlinks]{acronym}
\def\BibTeX{{\rm B\kern-.05em{\sc i\kern-.025em b}\kern-.08em
    T\kern-.1667em\lower.7ex\hbox{E}\kern-.125emX}}

\let\labelindent\relax
\usepackage[inline,shortlabels]{enumitem}

\usepackage{tikz}
\usepackage{pgfplots}
\pgfplotsset{compat=1.18}
\usepgfplotslibrary{statistics}
\usepgfplotslibrary{fillbetween} 
\usetikzlibrary{arrows}
\usetikzlibrary{decorations.pathreplacing,decorations.markings}
\usetikzlibrary{positioning}
\usetikzlibrary{calc}
\usetikzlibrary{fit}
\usetikzlibrary{patterns}
\usetikzlibrary{calc}
\usetikzlibrary{plotmarks}

\usepackage[americanvoltages]{circuitikz}
\ctikzset{bipoles/length=0.85cm}
\ctikzset{bipoles/thickness=1}

\tikzset{every picture/.style={line width=0.65pt}}

\definecolor{vgDarkBlue}{HTML}{00008B}
\definecolor{vgOrange}{HTML}{F7941D}
\definecolor{vgGreen}{HTML}{009B77}
\definecolor{vgLightBlue}{HTML}{1E90FF}
\definecolor{vgYellow}{HTML}{FFD700}
\definecolor{vgRed}{HTML}{FF0000}

\pgfplotsset{every axis/.append style={semithick,tick style={major tick
      length=4pt,semithick,black}}}

\pgfkeys{/pgfplots/x axis shift down/.style={
    x axis line style={yshift=-#1},
    xtick style={yshift=-#1},
    tick align=outside,
    xticklabel shift={#1}}}

\pgfkeys{/pgfplots/y axis shift left/.style={
    y axis line style={xshift=-#1},
    ytick style={xshift=-#1},
    yticklabel shift={#1},
    scaled y ticks = false,
    tick align=outside,
    y tick label style={/pgf/number format/fixed,
    }
  }
}

\pgfplotsset{myPlot/.style={%
    width=8cm,
    height=2.8cm,
    separate axis lines,
    axis x line*=bottom,
    x axis shift down = 7pt,
    enlarge x limits=false,
    axis y line*=left,
    y axis shift left = 10pt,
    enlarge y limits={abs=.25pt},
    enlarge x limits={abs=.25pt},
  }
}

\pgfplotsset{resultsPlot/.style={%
    myPlot,
    xtick = {0, 48, 96, 144, 192, 240, 288, 336},
    xticklabels = {$0$, $1$, $2$, $3$, $4$, $5$, $6$, $7$},
    xmin = 0,
    xmax = 336,
    clip=false,
  }
}

\newlength{\resultsPlotShift}
\pgfplotsset{resultsPlotStackedPower/.style={%
    resultsPlot,
    height = 3.5cm,
    stack plots=y, stack negative=separate, area style, enlarge x limits=false,
    xlabel = {},
    ylabel={Power $[\unit{pu}]$},
    y label style={yshift=+2em},
    ymin = -2,
    ymax = 2,
    xtick = {0, 48, ..., 336},
    xticklabels = {~},
    legend columns=4,
    legend style={
      at={(0.45, 1.0)},
      anchor=south,
      draw=none,
      fill=none,
      legend cell align=left,
      /tikz/every even column/.append style={column sep=0.3cm}
      },
  }
}

\pgfplotsset{resultsPlotStackedEnergy/.style={%
    resultsPlot,
    height = 2.5cm,
    yshift = -2.5cm+0.6cm,
    ymin = 0.5,
    ymax = 6.5,
    ytick = {0.5, 3.5, 6.5},
    ylabel={\begin{tabular}{c}  Energy \\ $[\unit{pu\,h}]$ \end{tabular}},
    y label style={yshift=+2em},
    xlabel = {},
    xtick = {0, 48, ..., 336},
    xticklabels = {~},
  }
}

\pgfplotsset{resultsPlotStackedLinePower/.style={%
    resultsPlot,
    height = 3cm,
    yshift = -2.5cm + 0.6cm - 3cm + 0.6cm,
    ymin = -1.3,
    ymax = 1.3,
    ytick = {-1.3, 0, 1.3},
    ylabel = {\begin{tabular}{c} Line \\ power $[\unit{pu}]$ \end{tabular}},
    y label style={yshift=+2em},
    xlabel = {Time $[\unit{d}]$},
    x label style={yshift=-1.5em},
  }
}

\newcommand{\PlotLimit}[1]{%
  \addplot[limit, forget plot] plot coordinates{
    (1,#1)
    (7*48,#1)
  };
}

\newcommand{\plotCaseStudyResultsFiveNodes}[1]{
  \begin{axis}[%
    resultsPlotStackedPower,
    ]
    \addplot [linestyleStoragePower stacked] table [x=t, y=storagePower, col sep=comma] {#1} \closedcycle;
    \addlegendentry{Storage}
    \addplot [linestyleRes stacked] table [x=t, y=renewablePower, col sep=comma] {#1} \closedcycle;
    \addlegendentry{Wind}
    \addplot [linestyleThermal stacked] table [x=t, y=thermalPower, col sep=comma] {#1} \closedcycle;
    \addlegendentry{Conv.}
    \addplot [linestyleDemand stacked] table [x=t, y=loadPower, col sep=comma] {#1} \closedcycle;
    \addlegendentry{Load}

    \addlegendimage{line legend, color = vgDarkBlue}
    \addlegendentry{Line 1}
    \addlegendimage{line legend, color = vgRed}
    \addlegendentry{Line 2}
    \addlegendimage{line legend, color = vgOrange}
    \addlegendentry{Line 3}
    \addlegendimage{line legend, color = vgGreen}
    \addlegendentry{Line 4}

  \end{axis}

  \begin{axis}[%
    resultsPlotStackedEnergy,
    ]

    \PlotLimit{6.5}
    \PlotLimit{0.5}

    \addplot[linestyleStorageEnergy] table [x=t, y=storedEnergy, col sep=comma]{#1};

    \ifthenelse{\equal{#1}{data/smallGrid/ideal.csv}}{}{
      \addplot[linestyleStorageEnergy, dashed] table [x=t, y=storedEnergy, col sep=comma]{data/smallGrid/ideal.csv};
      \draw[shorten >=6pt, shorten <=1.0pt, draw = gray] (axis cs:96,1.8919) -- (axis cs:66, 4.5) node[] {prescient};
    }
  \end{axis}

  \begin{axis}[%
    resultsPlotStackedLinePower,
    ]

    \PlotLimit{-1.3}
    \PlotLimit{1.3}

    \addplot[color = vgDarkBlue] table [x=t, y=pE1, col sep=comma]{#1};
    \addplot[color = vgRed] table [x=t, y=pE2, col sep=comma]{#1};
    \addplot[color = vgOrange] table [x=t, y=pE3, col sep=comma]{#1};
    \addplot[color = vgGreen] table [x=t, y=pE4, col sep=comma]{#1};

  \end{axis}
}

\newcommand{\plotCaseStudyResultsOreo}[2]{
  \begin{axis}[%
    resultsPlotStackedPower,
    ]
    \addplot [linestyleStoragePower stacked] table [x=t, y=storagePower1, col sep=comma] {#1} \closedcycle;
    \addlegendentry{Storage 1}
    \addplot [color=vgRed, fill=vgRed!20!white] table [x=t, y=storagePower2, col sep=comma] {#1} \closedcycle;
    \addlegendentry{Storage 2}
    \addplot [linestyleWind stacked] table [x=t, y=renewablePower1, col sep=comma] {#1} \closedcycle;
    \addlegendentry{RES 1}
    \addplot [linestylePv stacked] table [x=t, y=renewablePower2, col sep=comma] {#1} \closedcycle;
    \addlegendentry{RES 2}
    \addplot [linestyleThermal stacked] table [x=t, y=thermalPower1, col sep=comma] {#1} \closedcycle;
    \addlegendentry{Conv. 1}
    \addplot [color=black, fill=black!60] table [x=t, y=thermalPower2, col sep=comma] {#1} \closedcycle;
    \addlegendentry{Conv. 2}
    \addplot [linestyleDemand stacked] table [x=t, y=loadPower, col sep=comma] {#1} \closedcycle;
    \addlegendentry{Load}

    \ifthenelse{\equal{#1}{data/oreoGrid/ideal.csv}}{}{
      \addplot [linestyleStoragePower, dashed, line width = 1.5pt] table [x=t, y=storagePower1, col sep=comma] {data/oreoGrid/ideal.csv} \closedcycle;
    }

    \addlegendimage{line legend, color = vgDarkBlue}
    \addlegendentry{Line 1}
    \addlegendimage{line legend, color = vgRed}
    \addlegendentry{Line 2}
    \addlegendimage{line legend, color = vgOrange}
    \addlegendentry{Line 3}
    \addlegendimage{line legend, color = vgGreen}
    \addlegendentry{Line 4}
    \addlegendimage{line legend, color = vgLightBlue}
    \addlegendentry{Line 5}

  \end{axis}

  \begin{axis}[%
    resultsPlotStackedEnergy,
    ]

    \PlotLimit{6.5}
    \PlotLimit{3.5}
    \PlotLimit{0.5}

    \addplot[linestyleStorageEnergy] table [x=t, y=storedEnergy1, col sep=comma]{#1};
    \addplot[color=vgRed] table [x=t, y=storedEnergy2, col sep=comma]{#1};

    \ifthenelse{\equal{#1}{data/oreoGrid/ideal.csv}}{}{
      \addplot[linestyleStorageEnergy, dashed, line width=1.5pt] table [x=t, y=storedEnergy1, col sep=comma]{data/oreoGrid/ideal.csv};
      \addplot[color=vgRed, dashed] table [x=t, y=storedEnergy2, col sep=comma]{data/oreoGrid/ideal.csv};
      \draw[shorten >=11pt, shorten <=1pt, draw = gray] (axis cs:79, 3.1066) -- (axis cs:40, 4.5) node[] {prescient};
    }

  \end{axis}

  \begin{axis}[%
    resultsPlotStackedLinePower,
    ]

    \PlotLimit{-1.3}
    \PlotLimit{1.3}

    \addplot[color = vgDarkBlue] table [x=t, y=pE1, col sep=comma]{#1};
    \addplot[color = vgRed] table [x=t, y=pE2, col sep=comma]{#1};
    \addplot[color = vgOrange] table [x=t, y=pE3, col sep=comma]{#1};
    \addplot[color = vgGreen] table [x=t, y=pE4, col sep=comma]{#1};
    \addplot[color = vgLightBlue] table [x=t, y=pE5, col sep=comma]{#1};

  \end{axis}
}

\pgfplotsset{sidenotePlot/.style={%
    width = 3.3cm,
    height=2.8cm,
    separate axis lines,
    axis x line*=bottom,
    x axis shift down = 4pt,
    enlarge x limits=false,
    axis y line*=left,
    y axis shift left = 7pt,
    enlarge y limits={abs=.25pt},
    enlarge x limits={abs=.25pt},
    legend columns=1,
    legend style={
      at={(1.2, 1.1)},
      yshift = .6,
      anchor=south east,
      draw=none,
      fill=none,
      legend cell align=left,
    },
    clip=false,
  }
}

\pgfplotsset{tufteBoxplot/.style={%
    myPlot,
    clip=false,
    line width = 0.65pt,
    width= 8cm,
    hide y axis,
    x tick label style={
    /pgf/number format/.cd,
        fixed,
        precision=2,
    /tikz/.cd
    }
  }
}

\pgfplotsset{
      boxplot/draw/median/.code={%
        \draw[mark size=2pt,/pgfplots/boxplot/every median/.try]
          \pgfextra
          \pgftransformshift{
            \pgfplotsboxplotpointabbox
              {\pgfplotsboxplotvalue{median}}
              {0.5}
          }
          \pgfsetfillcolor{black}
          \pgfuseplotmark{*}
          \endpgfextra
        ;
      },
  }

\pgfplotsset{linestyle boxplot/.style={%
  boxplot = {%
    every box/.style={draw=none, fill=none},
    whisker extend=0,
  },
  mark=o,
  every mark/.append style={mark size=1.0pt}, draw=black,
  }
}

\pgfplotsset{variablesPlot/.style={%
    sidenotePlot,
    height=3.3cm,
    width = 3.0cm,
    xmin = 0,
    xmax = 1,
    xtick = {0, 1},
    xlabel = {},
    ymin= 0,
    ymax= 2,
    clip = false,
    hide y axis,
  }
}

\pgfplotsset{histVSnormpdfPlot/.style={%
  sidenotePlot,
  xlabel={error},
  xtick = {-0.5, -0.4, ..., 0.5},
  x tick label style={
    /pgf/number format/.cd,
    fixed,
    precision=2,
    /tikz/.cd
    },
  },
}

\pgfplotsset{
  /pgfplots/ybar legend/.style={
    /pgfplots/legend image code/.code={%
      \draw[##1, draw = none, /tikz/.cd,bar width=1pt, yshift=-0.2em, bar shift=6*\pgfplotbarwidth]
          plot coordinates {(0.5*\pgfplotbarwidth,0.3em) (2.5*\pgfplotbarwidth, 0.6em) (4.5*\pgfplotbarwidth,0.2em) (6.5*\pgfplotbarwidth,0.4em) (8.5*\pgfplotbarwidth,0.15em)};},
  }
}

\pgfplotsset{fcPlot/.style={%
    sidenotePlot,
    ylabel={Power $[\unit{pu}]$},
    xmin = 0,
    xmax = 24,
    xtick = {0, 12, 24},
    xlabel = {Time instant},
    xticklabels = {$k-12$, $k$, $k+12$},
  }
}

\pgfplotsset{problemDroopPlot/.style={%
    myPlot,
    clip=false,
    width = 3.0cm,
    height = 2.8cm,
    xtick = {0, 5, 10},
    restrict x to domain = 0:10,
    xmin = 0,
    xmax = 10,
    xlabel={Time $k$},
  }
}

\pgfplotsset{problemDroopThemal/.style={%
    problemDroopPlot,
    ytick = {0, 1},
    ymin=0,
    ymax=1,
    ylabel = {Conv. $[\unit{pu}]$},
    xshift = 7cm,
  }
}

\pgfplotsset{problemDroopWind/.style={%
    problemDroopPlot,
    ytick = {0, 1, 2},
    ymin=0,
    ymax=2,
    ylabel = {Wind $[\unit{pu}]$},
    xshift=3.5cm,
    legend columns=3,
    legend style={
      at={(1.4, 1.2)}, 
      yshift = .6,
      anchor=south,
      draw=none,
      fill=none,
      legend cell align=left,
      /tikz/every even column/.append style={column sep=0.3cm}
    },
  }
}

\pgfplotsset{problemDroopLoad/.style={%
    problemDroopPlot,
    ytick = {0, -1},
    ymin=-1,
    ymax=0,
    ylabel = {Load $[\unit{pu}]$},
  }
}

\pgfplotsset{problemDroopBattery/.style={%
    problemDroopPlot,
    ytick = {-1, 0, 1},
    ymin=-1,
    ymax=1,
    ylabel = {Storage $[\unit{pu}]$},
    xshift=10.5cm,
  }
}

\pgfplotsset{detailedVariablesPlot/.style={%
    myPlot,
    clip=false,
    clip mode=individual,
    width = 2.5cm,
    height = 5cm,
    xtick = {0, 1},
    restrict x to domain = 0:1,
    xmin = 0,
    xmax = 1,
    xlabel={Stage $j$},
    ylabel = {},
  }
}

\pgfplotsset{detailedVariablesPlot conventional/.style={%
    detailedVariablesPlot,
    ytick = {0, 1},
    ymin=-0.15,
    ymax=1,
    title = {Conv. gen.},
    xshift = 4.8cm,
  }
}

\pgfplotsset{detailedVariablesPlot wind/.style={%
    detailedVariablesPlot,
    ytick = {0, 1, 2},
    ymin=0,
    ymax=2,
    title = {Wind},
    xshift=2.4cm,
    legend columns=3,
    legend style={
      at={(1.4, 1.2)}, 
      yshift = .6,
      anchor=south,
      draw=none,
      fill=none,
      legend cell align=left,
      /tikz/every even column/.append style={column sep=0.3cm}
    },
  }
}

\pgfplotsset{detailedVariablesPlot load/.style={%
    detailedVariablesPlot,
    ytick = {0, -1},
    ymin=-1,
    ymax=0,
    title = {Load},
  }
}

\pgfplotsset{detailedVariablesPlot battery/.style={%
    detailedVariablesPlot,
    ytick = {-1, 0, 1},
    ymin=-1,
    ymax=1,
    title = {Storage},
    xshift=7.3cm,
  }
}

\pgfplotsset{exampleMpcPlot/.style={%
    myPlot,
    clip=false,
    width = 3.0cm,
    height = 2.8cm,
    xtick = {0, 4, 8, 12},
    xticklabels = {~},
    restrict x to domain = 0:12,
    xmin = 0,
    xmax = 12,
    legend style={
      at={(1.0, 1.0)},
      yshift = 0.6cm,
      anchor=south east,
      draw=none,
      fill=none,
      inner sep=0pt, outer sep=0pt,
      legend cell align=left,
    },
  }
}

\pgfplotsset{exampleMpcThemal/.style={%
    exampleMpcPlot,
    ytick = {0, 1},
    ymin=0,
    ymax=1,
    ylabel = {Conv. $[\unit{pu}]$},
    xshift = 7cm,
  }
}

\pgfplotsset{exampleMpcWind/.style={%
    exampleMpcPlot,
    ytick = {0, 1, 2},
    ymin=0,
    ymax=2,
    ylabel = {Wind $[\unit{pu}]$},
    xshift=3.5cm,
  }
}

\pgfplotsset{exampleMpcLoad/.style={%
    exampleMpcPlot,
    ytick = {0, -1},
    ymin=-1,
    ymax=0,
    ylabel = {Load $[\unit{pu}]$},
  }
}

\pgfplotsset{exampleMpcBattery/.style={%
    exampleMpcPlot,
    ytick = {-1, 0, 1},
    ymin=-1,
    ymax=1,
    ylabel = {Storage $[\unit{pu}]$},
    xshift=10.5cm,
  }
}

\pgfplotsset{exampleMpcEnergy/.style={%
    exampleMpcPlot,
    ytick = {0.5, 3.5, 6.5},
    ymin=0.5,
    ymax=6.5,
    ylabel={Energy $[\unit{pu\,h}]$},
    xshift=10.5cm,
  }
}

\pgfplotsset{avarSetPlot/.style={%
  line width = 0.65pt,
  width= 3.8cm,
  height = 3.8cm,
    view = {-25}{-25},
    axis lines = middle,
  xmin=0,
  xmax=1.3,
  xlabel={$\pi^{\prime}_1$},
  xlabel style={anchor=south},
  ymin=0,
  ymax=1.3,
  ylabel={$\pi^{\prime}_2$},
  ylabel style={anchor=east},
  zmin=0,
  zmax=1.2,
  zlabel={$\pi^{\prime}_3$},
  zlabel style={anchor=west},
  xtick = {0, 1},
  ytick = {0, 1},
  ztick = {0, 1},
  clip=false,
  }
}

\tikzset{mgScheme/.style={%
  >=latex,
  line width = 0.65pt,
  color=black,
  }
}

\tikzset{invS/.style={
  rectangle,
  fill=white,
  draw=black,
  inner sep=0.1pt
  }
}

\tikzset{midArrow/.style={
  draw,
  decoration={
    markings,
    mark=at position 0.5     with \arrow{stealth},%
  },
  postaction=decorate
  }
}

\tikzset{
  bidirArrow/.style 2 args={ 
    draw,
    decoration={
      markings,
      mark=at position 0.4 with {
        \node[above] {#1}; 
        \arrow{stealth};
      },
      mark=at position 0.6 with {
        \node[below] {#2}; 
        \arrowreversed{stealth};
      },
    },
    postaction=decorate
  }
}

\pgfplotsset{linestyleMeanValue/.style = {solid, color= vgLightBlue}}
\pgfplotsset{linestyleMc/.style = {solid, color= vgGreen, opacity={0.025}}}
\pgfplotsset{linestyleScenarioTree/.style = {solid, color=black, mark=*, opacity = {0.35}, mark size=0.5}}
\pgfplotsset{linestyleScenarioTreeLegend/.style = {linestyleScenarioTree, opacity = .7}}

\pgfplotsset{powerSetpoint/.style = {const plot, color=vgGreen, line width = 2pt}}
\pgfplotsset{uncurtailedRes/.style = {color = vgDarkBlue, line width = 2pt}}
\pgfplotsset{measuredPower/.style = {color = vgLightBlue, line width = 0.65pt}}
\pgfplotsset{energy/.style = {draw = black, line width = 0.65pt, mark=*, mark size=0.5}}
\pgfplotsset{measuredPowerBackground/.style = {color = white, line width = 1pt, forget plot}}

\pgfplotsset{limit/.style = {const plot, color=black, dash pattern=on 1pt off 3pt on 3pt off 3pt}}

\pgfplotsset{limit/.style = {const plot, color=black, dash pattern=on 1pt off 3pt on 3pt off 3pt}}

\colorlet{colorThermal}{vgDarkBlue}
\colorlet{colorStorage}{vgOrange}
\colorlet{colorRes}{vgGreen}
\colorlet{colorDemand}{vgLightBlue}
\colorlet{colorWind}{vgGreen}
\colorlet{colorPv}{vgYellow}

\pgfplotsset{linestyleThermal stacked/.style = {%
  color=colorThermal,
  fill=colorThermal!20!white,
}}

\pgfplotsset{linestyleStorageEnergy/.style = {
  color=colorStorage,
}}

\pgfplotsset{linestyleStoragePower stacked/.style = {%
  color=colorStorage,
  fill=colorStorage!20!white
}}

\pgfplotsset{linestyleRes stacked/.style = {%
  color=colorRes,
  fill=colorRes!20!white,
}}

\pgfplotsset{linestyleDemand stacked/.style = {%
  color=colorDemand,
  fill=colorDemand!20!white,
}}

\pgfplotsset{linestyleWind stacked/.style = {%
  color=colorWind,
  fill=colorWind!20!white,
}}

\pgfplotsset{linestylePv stacked/.style = {%
  color=colorPv,
  fill=colorPv!20!white,
}}

\usepgfplotslibrary{external}
\tikzset{external/system call={pdflatex \tikzexternalcheckshellescape -halt-on-error
    -interaction=batchmode -jobname "\image" "\texsource" &&
    pdf2ps -r1200 "\image".pdf "\image".ps &&
    ps2pdf -dAutoRotatePages=/None "\image".ps "\image".pdf}}

\tikzstyle{block} = [%
line width = 0.65pt,
draw, fill=none,
rectangle,
minimum height=2.5em,
minimum width=4.8em,
align=center,
inner sep=1mm,
]

\pgfplotsset{linestyle boxplot/.style={%
  boxplot = {%
    every box/.style={draw=none, fill=none},
    whisker extend=0,
    },
    mark=*,
    every mark/.append style={mark size=0.7pt, line width=0pt, opacity=0.6, fill=#1}, draw=#1,
    boxplot/draw/median/.code={%
          \draw[mark size=1.5pt, /pgfplots/boxplot/every median/.try]
          \pgfextra
          \pgftransformshift{
            \pgfplotsboxplotpointabbox
              {\pgfplotsboxplotvalue{median}}
              {0.5}
          }
          \pgfsetfillcolor{#1}
          \pgfuseplotmark{*}
          \endpgfextra
        ;
      },
  },
}

\theoremstyle{definition}
\newtheorem{definition}{Definition}

\newtheorem{problem}{Problem}
\newtheorem{remark}{Remark}
\newtheorem{assumption}{Assumption}
\newtheorem{theorem}{Theorem}

\acrodef{mpc}[MPC]{model predictive control}
\acrodef{res}[RES]{renewable energy sources}
\acrodef{ac}[AC]{alternating current}
\acrodef{dc}[DC]{direct current}
\acrodef{bim}[BIM]{bus injection model}
\acrodef{mg}[MG]{microgrid}
\acrodef{bess}[BESS]{battery energy storage system}
\acrodef{socp}[SOCP]{second-order cone programming}
\acrodef{lti}[LTI]{linear time-invariant}
\acrodef{milp}[MILP]{mixed-integer linear programming}
\acrodef{minlp}[MINLP]{mixed-integer nonlinear problem}
\acrodef{sdp}[SDP]{semidefinite programming}
\acrodef{misocp}[MISOCP]{mixed-integer second-order cone programming}
\acrodef{deepc}[DeePC]{data-enabled predictive control}
\acrodef{der}[DER]{distributed energy resources}
\acrodef{dcp}[DCP]{disciplined convex programming}
\acrodef{io}[I/O]{input/output}
\acrodef{pmu}[PMU]{phasor measurement unit}

\title{\LARGE \bf Data-Driven Active Power Flow Modeling: \\ A Behavioral Systems Approach}

\author{Sebastian Otzen$^{1}$, Hannes M. H. Wolf$^{2}$ and Christian A. Hans$^{2}$%
\thanks{$^{1}$S. Otzen is with Jouly, Berlin, Germany. {\tt\small sebastian@jouly.ai}}%
\thanks{$^{2}$H. M. H. Wolf and C. A. Hans are with the Automation and Sensorics in Networked Systems Group, University of Kassel, Germany. {\tt\small h.wolf@uni-kassel.de}, {\tt\small hans@uni-kassel.de}}%
}

\begin{document}

\maketitle
\thispagestyle{empty}
\pagestyle{empty}

\begin{abstract}
The increasing decentralization of power systems driven by a large number of renewable energy sources poses challenges in power flow optimization: Partially unknown power line properties can render model-based approaches unsuitable.
With the increasing deployment of sensors, data-driven methods rise as a promising alternative, offering flexibility to adapt changes and deal with unknown properties. 
In this paper, we propose a novel data-driven representation of nonlinear active power flow equations for radial grids based on Willems' Fundamental Lemma. 
Our approach allows for direct integration of input/output data into active power flow optimization, enabling cost minimization and constraint enforcement without requiring explicit knowledge of the electrical properties of the grid.
Moreover, we derive a computationally tractable convex relaxation and show in a numerical case study that our approaches yield results that are identical to optimal active power flow formulations with known parameters.

\end{abstract}



\section{Introduction}

Power systems are shifting from centralized generation to decentralized, small-scale units such as photovoltaic generators and wind turbines. 
This poses new challenges in the operation of power systems through an increase in the number of generation units and bidirectional power flows~\cite{Bordons2020}.
Special challenges arise in power flow optimization, which traditionally relies on known topologies and accurate models~\cite{Low2014}. 
Similarly, accurate power flow calculations have become increasingly important, as operators must respect equipment limits while integrating uncertain renewable sources~\cite{Yu2025}.

Meanwhile, the ongoing digitization of power grids increases the availability of high-resolution data, enabling novel methods. 
Existing approaches either (i) include an explicit modeling or system identification step or (ii) employ \ac{io} data right away to represent systems:

Examples for type (i) are \cite{Liu2019} and \cite{Guo2022}, where power flow is modeled using regression-based methods, mapping node measurements to power flows over power lines. 
While~\cite{Liu2019} uses linear mappings, \cite{Guo2022} makes use of Koopman operator theory, elevating node measurements into a higher dimensional space where a mapping to power flows is learned. 

Type (ii) is based on behavioral systems theory \cite{Polderman1998,Willems2004,Berberich2020} and eliminates the need for an explicit identification step. 
Recent studies \cite{Markovsky2023, Lipka2024} have employed such approaches for predictive control of power systems.
Moreover, in \cite{Bilgic2022}, a behavioral approach for controlling multi-energy systems was presented. 
The methodology incorporated uncertain forecasts of exogenous signals into the data-driven predictive control formulation. 
A similar approach was employed in~\cite{Molodchyk2025a} for a data-driven multi-stage \acs{dc} optimal power flow problem, which requires only measured data of nodal power injection and does not need information about the grid topology. 
However, \acs{dc} power flow does not account for transmission losses, rendering the approach less accurate.

To enable applications to nonlinear systems and enhance robustness to noise, the approach presented in~\cite{Coulson2019} introduces regularization into the optimal control problem. 
Moreover, the authors of \cite{Berberich2020} extend the application of behavioral systems theory to nonlinear Hammerstein and Wiener systems. 
However, to the best of our knowledge, such data-driven representations have not yet been employed for nonlinear power flow. 
We aim to fill this gap by making the following contributions.
\begin{enumerate*}[(i)]
    \item We derive a data-driven nonlinear active power flow representation for radial networks based on behavioral systems theory that can be used in optimization problems. 
    Our formulation only requires \ac{io} data of power lines, i.e., phase angle differences and transmitted power.
    \item We propose convex relaxations of aforementioned optimization problems, enabling the use of off-the-shelf solvers and convex optimization modeling frameworks.
    \item We demonstrate the applicability of the proposed approaches in a numerical case study on microgrid operation control.
\end{enumerate*}

The remainder of this paper is structured as follows. Section~\ref{sec:math_pres} introduces mathematical preliminaries. 
Sections~\ref{sec:physics_based_nonlinear} and~\ref{sec:data-driven-NL-PF} introduce physics-based and data-driven power flow optimization problems, respectively. 
In Section~\ref{sec:application_example} we use aforementioned formulations in a microgrid example. 
Section~\ref{sec:casestudy} presents a numerical case study on this example. Section~\ref{sec:conclusion} concludes the paper.


\section{Mathematical Preliminaries}\label{sec:math_pres}

In this section, we cover relevant notations and concepts from data-driven system representations and graph theory.

\subsection{Notations}

We denote real numbers, nonpositive real numbers, and nonnegative real numbers by $\mathbb{R}$, $\mathbb{R}_{\leq0}$ and $\mathbb{R}_{\geq0}$, respectively. 
The sets of positive and nonnegative integers are denoted $\mathbb{N}$ and $\mathbb{N}_0$, respectively. 
The set of Booleans is $\mathbb{B} = \{0,1\}$ and the set of complex numbers is $\mathbb{C}$. 
We express a complex number $x \in \mathbb{C}$ in polar form as $x = \hat{x} e^{\imath\theta}$, where $\hat{x} = |x| \in \mathbb{R}$ is the absolute value of $x$, $\theta \in \mathbb{R}$ is the angle of $x$, $e$ is Euler's number, and $\imath$ is the imaginary unit. 
In Cartesian form, $x$ can be expressed by $x = \mathrm{Re}(x) + \imath\,\mathrm{Im}(x)$. 

A sequence ${\mathbb{W} = \{w(a), w(a+1), \ldots, w(b)\}}$ is an ordered set of $b-a+1$ elements, where $a,b\in\mathbb{N}_0$, $b\geq a$ and $w(k) \in \mathbb{R}^{N_w}$ with discrete time instant $k\in\mathbb{N}_0$. 
Set $\mathbb{W}$ can be compactly written as $\{w(k)\}_{k=a}^b$.
Throughout this work, $[w]_{k=a}^b$ is shorthand for
$
    \begin{bmatrix}
        w(a)^T~
        w(a+1)^T~
        \cdots~
        w(b)^T
    \end{bmatrix}^T
$. 
Measured values are denoted by superscript $m$, e.g., $w^m(k)$.
Discrete-time instant $k$ refers to time $t = t_0 + k T_s$, where $t_0 \in \mathbb{R}_{\geq 0}$ is the initial time and $T_s \in \mathbb{R}_{\geq 0}$ the sampling time.
We use $w(k+h|k)$ to refer to a prediction for time instant $k+h,\, h\in\mathbb{N}_0$ that is performed at time $k$. 

\subsection{Hankel Matrix}

Consider a sequence $\mathbb{W} = \{w(k)\}_{k=0}^{N-1}$ of length $N \in \mathbb{N}$.
The associated Hankel matrix of order $L \in \mathbb{N}$, $L < N$, is
\begin{equation}\label{eq:hankel_matrices}
H_L(\mathbb{W}) =
\begin{bmatrix}
w(0) & w(1) & \cdots & w(N-L) \\
w(1) & w(2) & \cdots & w(N-L+1) \\
\vdots & \vdots & \ddots & \vdots \\
w(L-1) & w(L) & \cdots & w(N-1)
\end{bmatrix}\hspace{-0.3em}.
\end{equation}
To determine if a sequence allows us to reconstruct the behavior of a system, we recall the following definition.

\begin{definition}[Persistency of Excitation \cite{Willems2004}]\label{def:persistency_of_excitation}
A sequence $ \mathbb{W} = \{ w(k) \}_{k=0}^{N-1} $ with $ w(k) \in \mathbb{R}^{N_w}$ is said to be persistently exciting of order $L$ if $ H_L(\mathbb{W}) $ has full row rank, i.e., if
\begin{equation}
\text{rank}(H_L(\mathbb{W})) = N_w L.
\end{equation} 
\end{definition}

\subsection{Hammerstein Systems}

Consider matrices $A\in\mathbb{R}^{N_x\times N_x},{B}\in\mathbb{R}^{N_x\times \tilde{N}_u},C\in\mathbb{R}^{N_y\times N_x},{D}\in\mathbb{R}^{N_y\times \tilde{N}_u}$.
Moreover, consider state $x(k)\in\mathbb{R}^{N_x}$, input $u(k)\in\mathbb{R}^{N_u}$ and output $y(k)\in\mathbb{R}^{N_y}$. 
With them we can describe time-invariant Hammerstein systems as~\cite{Berberich2020}
\begin{subequations}
    \label{eq:hammerstein_model}
    \begin{align}
        x(k+1) &= A x(k) + {B} \nu(u(k)), \quad x(0) = x_0,  \\
    y(k) &= C x(k) + {D} \nu(u(k)),
    \end{align}
\end{subequations}
with the nonlinear mapping $\nu:\mathbb{R}^{N_u} \rightarrow \mathbb{R}^{\tilde{N}_u}$. 
This can be expressed by a linear combination of known basis functions $\phi_i: \mathbb{R}^{N_u} \rightarrow \mathbb{R}$, $i\in[1,r]\subset\mathbb{N}$ whose span includes $\nu$, i.e.,
\begin{align}
    \nu(u(k))= \textstyle\sum_{i=1}^r \zeta_{i}\phi_i(u(k)) = \zeta \phi(u(k)),
\end{align}
with $\zeta = [\zeta_i]_{i=1}^r \in \mathbb{R}^{\tilde{N}_u \times r}$, $\zeta_{i}\in\mathbb{R}^{\tilde{N}_u}$ not all zero, and
\begin{align}
    \label{eq:auxiliary_input_vector}
    \phi(u(k))= \begin{bmatrix} 
        \phi_1(u(k))~ 
        \cdots~
        \phi_{r}(u(k))
    \end{bmatrix}^T \in \mathbb{R}^r.
\end{align}
Combining $\zeta$ with $B$ and $D$ allows us to reformulate \eqref{eq:hammerstein_model} as 
\begin{subequations}
    \label{eq:hammerstein_model_aux}
    \begin{align}
        x(k+1) &= A x(k) + \tilde{B} \phi(u(k)), \quad x(0) = x_0,  \\
    y(k) &= C x(k) + \tilde{D} \phi(u(k)),
    \end{align}
\end{subequations}
with $\tilde{B} = B\zeta\in\mathbb{R}^{N_x\times r}$ and $\tilde{D}=D\zeta\in\mathbb{R}^{N_y\times r}$.
In what follows, we will use $\phi(k)$ as shorthand notation for $\phi(u(k))$.

\subsection{Data-Driven Representation of Hammerstein Systems}

\begin{theorem}[Data-driven Hammerstein System\cite{Berberich2020}]\label{hammerstein_fundamental_lema}
    Consider a measured \ac{io} trajectory $\{u^m(k),y^m(k)\}_{k=n}^{n+N-1}$ of a Hammerstein system \eqref{eq:hammerstein_model}. 
    Given that the associated auxiliary input $\{\phi^m(k)\}_{k=n}^{n+N-1}$, constructed via \eqref{eq:auxiliary_input_vector} using $\{u^m(k)\}_{k=n}^{n+N-1}$, is persistently exciting of order $L+N_x$, then, $\textstyle{\{u(k),y(k)\}_{k=l}^{l + L-1}}$ is a trajectory of \eqref{eq:hammerstein_model} if and only if there exists an $ \alpha \in \mathbb{R}^{N - L + 1} $ such that
\begin{subequations}    
\begin{equation} \label{eq:HL_system}
\begin{bmatrix}
    H_L(\{\phi^m(k)\}_{k=n}^{n+N-1}) \\
    H_L(\{y^m(k)\}_{k=n}^{n+N-1})
\end{bmatrix} \alpha = 
\begin{bmatrix}
    [\phi(k)]_{k=l}^{l+L-1} \\
    [y(k)]_{k=l}^{l+L-1}
\end{bmatrix},
\end{equation}
\end{subequations}
\end{theorem}
\begin{remark}
    Note that for Theorem~\ref{hammerstein_fundamental_lema}, the auxiliary input must contain the true basis functions. 
    In practical settings, sufficiently many basis functions or even kernel functions can be used to approximate the true ones~\cite{Berberich2020}. 
\end{remark}

\subsection{Graph Theory}

Following graph-theoretic terminology~\cite{Diestel2017}, we define an undirected weighted connected graph $\mathcal{G} = (\mathbb{V}, \mathbb{E}, y)$ through a set $\mathbb{V} \subseteq \mathbb{N}_0$ of $N_b \in \mathbb{N}$ nodes and $\mathbb{E} \subseteq [\mathbb{V}]^2$ of $N_e \in \mathbb{N}_0$ edges, where $[\mathbb{V}]^2$ is the set of all subsets of $\mathbb{V}$ with two elements.
The weighting function is $y: \mathbb{E} \rightarrow \mathbb{C}^3$ with $ y(\{i,j\}) = (y_{ij}, \bar{y}_{ij}, \bar{y}_{ji}) $.
Nodes $ i $ and $ j $ are adjacent if $ \{i,j\} \in \mathbb{E} $.
The set of adjacent nodes of $i \in \mathbb{V} $ is $ \mathbb{V}_i = \{ j \in \mathbb{V} \mid \{i, j\} \in \mathbb{E} \}$.
Let $\mathbb{E}=\{e_1,e_2,\dots,e_{N_e}\}$ be ordered such that for $e_l=\{i_l,j_l\}$ and $e_{l+1}=\{i_{l+1},j_{l+1}\}$, $l\in[1,N_e-1]\subset \mathbb{N}$, it holds that $i_l\leq i_{l+1}$ and $j_l\leq j_{l+1}$ if $i_l = i_{l+1}$.
Moreover, for $z_{e_l}\in\mathbb{R}^{N_z}$, we use $[z_{e_l}]_{\{e_l\}\in\mathbb{E}}$ as shorthand for $[z_{e_1}^T~z_{e_2}^T~\cdots~z_{e_{N_e}}^T]^T$.


\section{Physics-Based Power Flow}
\label{sec:physics_based_nonlinear}

In this section, we introduce the physics-based active power flow model and define an optimal power flow problem.
We model the electrical grid as a weighted undirected connected graph $\mathcal{G}$, where each edge $\{i,j\}\in\mathbb{E}$ represents a power line.
We start by posing some assumptions.

\begin{assumption}
    The power grid is balanced and symmetrical.
\end{assumption}

\begin{assumption}\label{ass:constant-voltages}
    The voltages at all buses are constant, i.e., $\hat{v}_i(k)=\hat{v}_i$ for all $i\in\mathbb{V}$. 
\end{assumption}

\begin{figure}[b]
    \centering

\tikzset{external/export next=true}
\begin{tikzpicture}[xscale=1.5, yscale = 1.0, font=\footnotesize, line width=0.7pt]
	
	\path[draw=none, fill=none] (0, 0) rectangle (1, 2);

	\draw
		(0, 0) to[open, v^<={$v_i e^{\imath\theta_i}$}] (0, 1.5)
		(0, 0) to[short, o-] (1, 0) to[short, -] (4, 0) to[short, -o] (5, 0)
		(0, 1.5) to[short, o-, i^>={$p_{e,ij}$}] (1, 1.5)
		(1, 0) to[generic, l_=$\bar{y}_{ij}$, *-*] (1, 1.5)
		(1, 1.5) to[short] (2.2, 1.5) to[generic, l_=$y_{ij}$] (2.8, 1.5) to[short] (4, 1.5)
		(4, 0) to[generic, l^=$\bar{y}_{ji}$, *-*] (4, 1.5)
		(5, 1.5) to[short, o-, i_>={$p_{e,ji}$}] (4, 1.5)
		(5, 0) to[open, v<={$v_j e^{\imath\theta_j}$}] (5, 1.5)
		;

\end{tikzpicture}
    \caption{$\pi$-equivalent circuit of a power line.}\label{fig:power_line}
\end{figure}

\begin{assumption}\label{ass:short-medium-lines}
    We assume short to medium long transmission lines up to a length of \enquote{about $\qty{200}{km}$} \cite{Kundur2007}.
    Thus, we can model them using the $\pi$-equivalent circuit in Figure~\ref{fig:power_line}.
\end{assumption}

\subsection{Nonlinear Power Flow Model}

Building on Assumption~\ref{ass:short-medium-lines}, we model connections between two nodes, i.e., edges in $\mathbb{E}$, using $\pi$-equivalent circuits.
The weighting function $y$ provides series admittance $y_{ij} = y_{ji} = g_{ij} + \imath b_{ij} $, with conductance $g_{ij}\in\mathbb{R}_{\geq0}$ and susceptance $b_{ij}\in\mathbb{R}$, and shunt admittances $\bar{y}_{ij} = \bar{g}_{ij} + \imath \bar{b}_{ij}$, $\bar{y}_{ji} = \bar{g}_{ji} + \imath \bar{b}_{ji}$.

\begin{remark}
    We can also represent more difficult connections between two nodes with our model, e.g., combinations of power lines, transformers and passive loads, since these can be transformed into $\pi$-equivalent using Kron-reduction \cite{dorfler_kron_2013}.
\end{remark}

Consider voltage phasors $v_i = \hat{v}_i e^{\imath \theta_i} \in \mathbb{C}$ at all nodes $i \in \mathbb{V}$.
At time $k$, the active power flow from node $i$ to $j$ is
\begin{multline} \label{eq:nonlinearPFsinglelineP}
    p_{e,ij}(k) = (\bar{g}_{ij}+g_{ij}) \hat{v}_i^2(k)   \\
              - \hat{v}_i(k) \hat{v}_j(k)\left( g_{ij} \cos(\theta_{ij}(k)) + b_{ij} \sin(\theta_{ij}(k)) \right),
\end{multline} 
with phase angle difference $\theta_{ij} = \theta_i - \theta_j$.
Assumption~\ref{ass:constant-voltages} allows us to rewrite \eqref{eq:nonlinearPFsinglelineP} as
\begin{align} 
    \label{eq:nonlinearPFsinglelineP_tilde}
    p_{e,ij}(k) &=  \breve{g}_{ij} - (\tilde{g}_{ij} \cos(\theta_{ij}(k)) + \tilde{b}_{ij} \sin(\theta_{ij}(k))),
\end{align} 
with
$\breve{g}_{ij}=(\bar{g}_{ij}+g_{ij})\hat{v}_i^2$,
$\tilde{g}_{ij}=g_{ij}\hat{v}_i\hat{v}_j$, and
$\tilde{b}_{ij}=b_{ij}\hat{v}_i\hat{v}_j$.

The net power injection $p_{g,i}\in \mathbb{R}$, i.e., the total power provided or consumed at node $i$, is equal to the sum of the power flowing into the connected power lines, i.e.,
\begin{align}
p_{g,i}(k) &= \textstyle\sum_{j\in \mathbb{V}_i} p_{e,ij}(k). \label{eq:line_to_bus_nonlinear}
\end{align}

\subsection{Nonlinear Optimal Power Flow}

We consider two active power flow values for each power line that we collect in $ p_e(k)= [p_{e,ij}(k) ~ p_{e,ji}(k)]^T_{\{i,j\}\in\mathbb{E}}$. 
Furthermore, we collect nodal power injections in $p_g(k) = [p_{g,i}(k)]_{i\in\mathbb{V}}$.
For a power system with $N_e$ power lines and $N_b$ nodes we collect line power flows and nodal power injections in $p(k) = [p_e(k)^T ~ p_g(k)^T]^T \in \mathbb{R}^{2N_e+N_b}$ and phase angle differences in $\theta(k)=[\theta_{ij}(k)]_{\{i,j\}\in\mathbb{E}} \in \mathbb{R}^{N_e}$.
Let us now employ \eqref{eq:nonlinearPFsinglelineP_tilde} and~\eqref{eq:line_to_bus_nonlinear} in an optimization problem.

\begin{problem}[Nonlinear optimal power flow]\label{prob:nonlinear_generic_opt}
\begin{align*}
    \min_{\substack{\theta(k),p(k)}} f(p(k))
\end{align*} 
subject to~\eqref{eq:line_to_bus_nonlinear} $\forall i\in\mathbb{V}$ and
\begin{subequations}
\begin{align*}
    p_{e,ij}(k) &=  \breve{g}_{ij} - (\tilde{g}_{ij} \cos(\theta_{ij}(k)) + \tilde{b}_{ij} \sin(\theta_{ij}(k))), \\
    p_{e,ji}(k) &=  \breve{g}_{ji} - (\tilde{g}_{ij} \cos(-\theta_{ij}(k)) + \tilde{b}_{ij} \sin(-\theta_{ij}(k))),
\end{align*}
\end{subequations}
 $\forall\{i,j\}\in\mathbb{E}$ as well as application-dependent constraints
\begin{subequations} \label{eq:application_constraints}
\begin{align}
    0 &\leq f_{\text{ineq}}\left(\theta(k),p(k)\right), \label{eq:inequality_constraints} \\
    0 &= f_{\text{eq}}\left(\theta(k),p(k)\right). \label{eq:equality_constraints}
\end{align}
\end{subequations}
\end{problem}

Here, the objective $f$ describes the cost of operation, e.g., in the form of generation cost or the transmission losses
\begin{align}
    p_{loss}(k) = \textstyle\sum_{i \in \mathbb{V}} p_{g, i}(k). \label{eq:transmission_losses}
\end{align}


\section{Data-Driven Power Flow}
\label{sec:data-driven-NL-PF}

In what follows, we transform the physics-based model into a Hammerstein system using appropriate basis functions. 
We then apply Theorem~\ref{hammerstein_fundamental_lema} and employ the resulting data-driven active power flow into an optimization problem.
We again start with some assumptions.

\begin{assumption}\label{ass:radial}
    We consider radial networks, which are widely present in microgrid settings~\cite{Lasseter}.
    The radial structure ensures unique power flow solutions and enables exact convex relaxations of associated formulations (see Remark~\ref{rem:convex_relaxation}).
\end{assumption}

\begin{assumption}
	Voltage phase angles $\theta_{ij}(k)$ are available for all lines, e.g., through time-synchronized \acp{pmu} or micro-\acp{pmu}.
	Moreover, we assume that active power measurements of all lines are available.
\end{assumption}

\subsection{Data-Driven Representation of Nonlinear Power Flow}

Equation \eqref{eq:nonlinearPFsinglelineP_tilde} for $\{i,j\}\in\mathbb{E}$ can be understood as a stateless \ac{io} system with input $u(k)=\theta_{ij}(k)$ and output $y(k)=p_{e,ij}(k)$. 
We can transform it into a Hammerstein system~\eqref{eq:hammerstein_model_aux} using the basis functions $1$, $\cos(\theta_{ij}(k))$, and $\sin(\theta_{ij}(k))$, i.e., 
\begin{align}\label{eq:basis_functions_ij}
    \phi(k) & = \begin{bmatrix}
    1~\cos(\theta_{ij}(k))~\sin(\theta_{ij}(k))
\end{bmatrix}^T. 
\end{align}
This allows us to express \eqref{eq:nonlinearPFsinglelineP_tilde} as
\begin{align}
    p_{e,ij}(k) &=  [\breve{g}_{ij} \quad-\tilde{g}_{ij} \quad-\tilde{b}_{ij}]\phi(k) \label{eq:hammersteinSingleLineP}.
\end{align}
By virtue of Theorem~\ref{hammerstein_fundamental_lema}, this can be represented in a data-driven manner as
\begin{equation}\label{eq:data-driven_powerflow_representation}
\begin{bmatrix}
    H_1(\{\phi^m(k)\}_{k=n}^{n+N-1}) \\
    H_1(\{p_{e,ij}^m(k)\}_{k=n}^{n+N-1})
\end{bmatrix} \alpha(k) = 
\begin{bmatrix}
    {\phi}(k) \\
    {p}_{e,ij}(k)
\end{bmatrix}.
\end{equation}

\begin{remark}
Note that we "consider algebraic relations as differential equations of order zero"~\cite[p.~12]{Polderman1998}, i.e., $ N_x = 0 $, $ A = 0 $, $ \tilde{B} = 0 $, $ C = 0 $, and $ \tilde{D} \neq 0 $. 
Consequently, $ L = 1 $. 
\end{remark}

\subsection{Data-Driven Optimal Power Flow}\label{sec:dd_opf_generic}
To employ \eqref{eq:data-driven_powerflow_representation} in optimization problems,
we extend \eqref{eq:basis_functions_ij} by trigonometric basis functions for all power lines, i.e., 
\begin{align}\label{eq:data-driven-aux}
\phi(k) = \big[
        1 \quad [\cos(\theta_{ij}(k)) \quad \sin(\theta_{ij}(k))]_{\{i,j\}\in\mathbb{E}}^T
    \big]^T.
\end{align}
This allows us to derive a data-driven representation of the nonlinear active power flow for the entire grid using $\phi^m(k)$ which is constructed along the lines of~\eqref{eq:data-driven-aux} and $p_e^m(k)$.

\begin{remark}
Due to the symmetry properties, $\sin(\theta_{ji}(k))$ and ${\cos(\theta_{ji}(k))}$ do not need to be explicitly included.
\end{remark}

In optimization problems, we treat $\phi(k)$ as a vector of decision variables. 
To maintain the inherent relation between the trigonometric components, we enforce ${\cos^2(\theta_{ij}(k)) + \sin^2(\theta_{ij}(k)) = 1}$ for all $\{i,j\}\in\mathbb{E}$, i.e., $\phi_l^2(k) + \phi_{l+1}^2(k) = 1$ for $l \in \{2, 4, \ldots,2N_e \}$. 
With $\alpha(k) \in \mathbb{R}^{N}$, we formulate the following optimization problem.

\begin{problem}[Data-driven optimal power flow]\label{prob:dd_generic_opt}
\begin{align*}
    \min_{\substack{\alpha(k),\phi(k),p(k)}} f(p(k))
\end{align*}
subject to \eqref{eq:line_to_bus_nonlinear} $\forall i\in\mathbb{V}$, \eqref{eq:application_constraints} and
\begin{subequations}
\begin{equation}\label{eq:data-driven_powerflow_constraint_generic}
\begin{bmatrix}
    H_1(\{\phi^m(k)\}_{k=n}^{n+N-1}) \\
    H_1(\{p_e^m(k)\}_{k=n}^{n+N-1})
\end{bmatrix} \alpha(k) = 
\begin{bmatrix}
    {\phi}(k) \\
    {p}_e(k)
\end{bmatrix},
\end{equation}
\begin{equation}\label{eq:nonconvex_constraint_generic}
    \phi_l^2(k) + \phi_{l+1}^2(k) = 1, \quad \forall l \in \{2,4, \ldots,2N_e\}.
\end{equation}
\end{subequations}
\end{problem}

  \begin{remark}\label{rem:scalability}     
      For a radial network with $N_e = N_b - 1$ lines and $N_b$ nodes, Problem~\ref{prob:dd_generic_opt} introduces $\phi(k) \in \mathbb{R}^{2N_e+1}$ and $\alpha(k) \in \mathbb{R}^N$ as decision variables.
      Since Definition~\ref{def:persistency_of_excitation} requires $N \geq 2N_e + 1 = 2N_b - 1$, the total number of decision variables and constraints grows linearly in $N_b$.                        
  \end{remark}   

Due to~\eqref{eq:nonconvex_constraint_generic}, Problem~\ref{prob:dd_generic_opt} is nonconvex. 
Such problems are computationally demanding and finding a global optimum can be hard. 
This issue is addressed in the next section.

\subsection{Convex Data-Driven Optimal Power Flow}\label{sec:convex_opf}

Convex formulations guarantee convergence to a globally optimal solution and improved scalability.
Inspired by~\cite{Jabr2006}, we therefore propose a convex relaxation that replaces \eqref{eq:nonconvex_constraint_generic} with the unit disk~\eqref{eq:convex_constraint_generic}.
To promote solutions on the unit disk boundary, we add a regularization term that maximizes cosine basis functions.
The problem reads as follows.

\begin{problem}[Convex data-driven optimal power flow]\label{prob:convex_dd_mpc_acopf_generic}
\begin{align*}
    \min_{\substack{\alpha(k),\phi(k),p(k)}} f(p(k)) - \beta \textstyle\sum_{i=1}^{N_e} \phi_{2i}(k)
\end{align*}
subject to~\eqref{eq:line_to_bus_nonlinear} $\forall i\in\mathbb{V}$, \eqref{eq:application_constraints}, \eqref{eq:data-driven_powerflow_constraint_generic} and
\begin{equation}\label{eq:convex_constraint_generic}
    \phi_l^2(k) + \phi_{l+1}^2(k) \leq 1, \quad \forall l \in  \{2,4, \ldots,2N_e\}.
\end{equation}
\end{problem}

\begin{remark}\label{rem:convex_relaxation}
For radial networks, replacing~\eqref{eq:nonconvex_constraint_generic} by~\eqref{eq:convex_constraint_generic} is exact if (C1) the cost function $f(p)$ is strictly increasing in all nodal power injections and (C2) the phase angle differences are sufficiently small \cite[Theorem~8]{Low2014a}.
In light of Assumption~\ref{ass:short-medium-lines}, (C2) is typically satisfied.
If (C1) holds, $\beta$ can be set to zero.
Otherwise, a small $\beta > 0$ promotes solutions on the unit circle.
The parameter $\beta$ should be chosen such that the resulting cost distortion $\beta N_e\theta^2_{\max}/2$, where $\theta_{\max}$ is the largest phase angle difference, remains negligible.
\end{remark}


\section{Microgrid Application Example}
\label{sec:application_example}

\begin{figure}
    \centering

\tikzset{external/export next=true}

\input{figures/NewNodeCommands.tex}

\newcommand{\wInPlot}[2]{}
\newcommand{\pInPlot}[2]{$p_{s,#1}$}
\newcommand{\pEPlot}[1]{$p_{e,#1}$}

\newcommand{\wDPlot}[1]{$w_{d,#1}$}
\newcommand{\wRPlot}[1]{$w_{r,#1}$}

\newcommand{\pTPlot}[1]{$p_{t,#1}$}
\newcommand{\pSPlot}[1]{$p_{s,#1}$}
\newcommand{\pRPlot}[1]{$p_{r,#1}$}
\newcommand{\pDPlot}[1]{$p_{d,#1}$}

\tikzset{
  leftRightArrow/.style={
    draw,
    decoration={
      markings,
      mark=at position 0.25 with {\arrow{stealth}},
      mark=at position 0.75 with {\arrowreversed{stealth}}
    },
    postaction=decorate
  }
}

\tikzset{
  leftRightArrow2/.style={
    draw,
    decoration={
      markings,
      mark=at position 0.3 with {\arrow{stealth}},
      mark=at position 0.65 with {\arrowreversed{stealth}}
    },
    postaction=decorate
  }
}

\begin{tikzpicture}[
	mgScheme,
	scale=1.1, 
	font = \footnotesize,
	powerLine/.style={draw},
]


\path (-2.5, 1) node[invS] (1) {\GraphTristateT};

\path[midArrow] (1) -- node[below]{\pTPlot{1}} ++(1.0, 0) coordinate (bar1);
\path[powerLine] (bar1)++(0, 0.25) node[yshift = 3] {\tiny 1}  -- ++(0, -0.5);
\node[left = -0.5mm of 1] {Conv. 1};

\path (-2.5, -1) coordinate (2);
\path (2) ++ (1, 0) coordinate (bar2);
\path (2) ++ (0, 0.4) node[invS] (2s) {\GraphTristateS};
\node[anchor=north west, inner sep=1pt, font=\footnotesize] at (2s.north west) {$x_1$};
\node[left = -0.5mm of 2s] {BESS 1};
\path (2) ++ (0, -0.4) node[invS] (2r) {\GraphTristateR};

\path[midArrow] (2s) -- node[below]{\pSPlot{1}} ++(1.0, 0) coordinate (bar2s);
\path[midArrow] (2r) -- node[below]{\pRPlot{1}} ++(1.0, 0) coordinate (bar2r);
\node[below = -0.5mm of 2r] {\wRPlot{1}};
\node[left = -0.5mm of 2r] {RES 1};

\path[powerLine] (bar2s) ++(0, 0.1) node[yshift = 3] {\tiny 2} -- (bar2r) -- ++(0, -0.1);

\path (2.5, -1.2) node[invS] (3) {\GraphTristateT};

\path[midArrow] (3) -- node[below]{\pTPlot{2}} ++(-1, 0) coordinate (bar3);
\path[powerLine] (bar3)++(0, 0.25)node[yshift = 3] {\tiny 3} -- ++(0, -0.5);
\node[right = -0.5mm of 3] {Conv. 2};

\path (2.5, 0.6) coordinate (4);
\path (4) ++ (-1.0, 0) coordinate (bar4);
\path (4) ++ (0, 0.4) node[invS] (4s) {\GraphTristateS} ;
\node[anchor=north west, inner sep=1pt, font=\footnotesize] at (4s.north west) {$x_2$};
\node[right = -0.5mm of 4s] {BESS 2};

\path (4) ++ (0, -0.4) node[invS] (4r) {\GraphTristatePV};
\node[below = -0.5mm of 4r] {\wRPlot{2}};
\node[right = -0.5mm of 4r] {RES 2};

\path[midArrow] (4s) -- node[below]{\pSPlot{2}} ++(-1.0, 0) coordinate (bar4s);
\path[midArrow] (4r) -- node[below]{\pRPlot{2}} ++(-1.0, 0) coordinate (bar4r);
\path[powerLine] (bar4s) ++(0, 0.1) node[yshift = 3] {\tiny 4}  -- (bar4r) -- ++(0, -0.1);

\path (0, -2.0) coordinate (5);
\path (5) ++(0, 0.5) coordinate (bar5) {};
\path[powerLine] (bar5)++(0.3, 0) -- ++(-0.6, 0) node[xshift = -3] {\tiny 5};
\path[powerLine, ->] (bar5) -- ++(0, -0.7) node[right]{\wDPlot{1}};
\path[midArrow] (bar5)++(0, -0.3) -- node[right]{\pDPlot{1}} ++(0, 0.001);

\path (bar1) ++ (0.0, 0.0) coordinate (l1bar1);
\path (l1bar1) ++ (0.2, 0.0) coordinate (l1bar1out);
\path (bar2) ++ (0.0, 0.3) coordinate (l1bar2);
\path (l1bar2) ++ (0.2, 0.0) coordinate (l1bar2out);
\path[leftRightArrow] (l1bar1) -- (l1bar1out) -- node[right]{} (l1bar2out) 
node[left, pos=0.2, rotate=90, anchor=north] {$p_{e,12}$}
node[left, pos=0.8, rotate=90, anchor=north] {$p_{e,21}$}
-- (l1bar2);

\path (bar3) ++ (0.0, 0.1) coordinate (l2bar3);
\path (l2bar3) ++ (-0.2, 0.0) coordinate (l2bar3out);

\path (bar2) ++ (0.0, -0.3) coordinate (l3bar2);
\path (bar5) ++ (-0.2, 0.0) coordinate (l3bar5);
\path (l3bar5) ++ (0.0, 0.2) coordinate (l3bar5out);
\path[leftRightArrow2] (l3bar2) -- 
node[above]{} (l3bar5out) 
node[left, pos=0.3, rotate=0, anchor=north] {$p_{e,25}$}
node[right, pos=0.8, rotate=0, anchor=south] {$p_{e,52}$}
-- (l3bar5);

\path (bar3) ++ (0.0, -0.1) coordinate (l4bar3);
\path (bar5) ++ (0.2, 0.0) coordinate (l4bar5);
\path (l4bar5) ++ (0.0, 0.2) coordinate (l4bar5out);
\path[leftRightArrow2] (l4bar3) -- 
node[above]{} (l4bar5out)
node[left, pos=0.3, rotate=0, anchor=south] {$p_{e,35}$}
node[right, pos=0.8, rotate=0, anchor=south] {$p_{e,53}$}
 -- (l4bar5);

\path(bar2) ++ (0.3, 0.0) coordinate (l5bar2out);
\path (bar4) ++ (0.0, 0.3) coordinate (l5bar4);
\path (l5bar4) ++ (-0.3, 0.0) coordinate (l5bar4out);
\path[leftRightArrow] (bar2) -- (l5bar2out) -- 
node[below, xshift = 5]{} (l5bar4out)
node[left, pos=0.17, sloped, anchor=north] {$p_{e,24}$}
node[right, pos=0.83, sloped, anchor=south] {$p_{e,42}$}
 -- (l5bar4);

\end{tikzpicture}
    \vspace{-1.5em}
    \caption{Example microgrid.}
    \label{fig:microgrid}
\end{figure}

In this section, we present a \ac{mg} application example that uses the proposed methods in a \ac{mpc} scheme.
While our approach is applicable to large radial networks (see Remark~\ref{rem:scalability}), we use an \ac{mg} due to its illustrative character.
The employed setup in Figure~\ref{fig:microgrid} is based on~\cite{Hans2021} and composed of $5$ buses $\mathbb{V} = \{1, 2, 3, 4, 5\}$ as well as $4$ power lines, i.e., $\mathbb{E} = \{\{1,2\}, \{2,4\}, \{2,5\}, \{3,5\}\}$.  
Furthermore, it includes two conventional units, two \acp{bess}, two \ac{res} and a load. 

\subsection{Constraints}

We collect the output power of the conventional generators in $p_t(k)=[p_{t,1}(k)~p_{t,2}(k)]^T\in\mathbb{R}^2_{\geq0}$ and their switch condition in  $\delta_{t}(k) = [\delta_{t,1}(k)~\delta_{t,2}(k)]^T\in\mathbb{B}^2$ where $\delta_{t,i}(k)=1$ means that generator $i\in\{1,2\}$ is enabled and $\delta_{t,i}(k)=0$ that it is disabled.
When enabled, the generators operate within their power limits, $p_t^{\min},~p_t^{\max} \in \mathbb{R}^{2}_{\geq 0}$, when disabled their respective power is zero.
This is captured by
\begin{align}\label{eq:p_t_ineq}
\text{diag}\left(p_t^{\min}\right)\delta_t(k) &\leq p_t(k) \leq \text{diag}\left(p_t^{\max}\right)\delta_t(k).
\end{align}

The \acp{bess} can either draw power from the grid or supply power to it.
The \acp{bess}' power $p_s=[p_{s,1}~p_{s,2}]^T\in\mathbb{R}^2$ must remain within the limits $p_s^{\min} \in \mathbb{R}^2_{\leq0}$, $p_s^{\max} \in \mathbb{R}^2_{\geq0}$, i.e.,
\begin{align}
p_s^{\min} &\leq p_s(k) \leq p_s^{\max}.
\end{align}
The stored energy of the \acp{bess} $x \in \mathbb{R}_{\geq 0}^{2}$ evolves over time based on $p_s$ and the initially stored energy $x_0 \in \mathbb{R}_{\geq 0}^{2}$, i.e.,
\begin{align}
    x(k+1) = A_s x(k) + B_sp_s(k), \quad x(0) = x_0,
\end{align}
where $A_s\in\mathbb{R}^{2\times 2}$ and $B_s\in\mathbb{R}^{2\times 2}$ are diagonal matrices.
The stored energy is bounded by $x^{\min}, \, x^{\max}\in \mathbb{R}_{\geq 0}^{2}$, i.e.,
\begin{align}
    x^{\min} \leq x(k) \leq x^{\max}.
\end{align}

\ac{res} provide power depending on wind or sunlight. 
Their infeed $p_{r}(k) =[p_{r,1}(k)~p_{r,2}(k)]^T\in\mathbb{R}^2_{\geq0}$ can not exceed the weather-dependent available power, $w_{r}(k) \in \mathbb{R}_{\geq 0}^{2}$, i.e.,
\begin{align}\label{eq:res_constraint}
    0 &\leq p_r(k) \leq w_r(k).
\end{align}

The active power flowing through power lines is limited by $ p_e^{\text{min}} \in \mathbb{R}^{8}_{\leq 0} $ and $ p_e^{\text{max}} \in \mathbb{R}^{8}_{\geq 0} $, i.e.,
\begin{align}\label{eq:active_powerflow_constraint}
    p_e^{\text{min}} \leq p_e(k) \leq p_e^{\text{max}}.
\end{align}
The load demand is $p_d(k) \in \mathbb{R}_{\leq 0}$.
The consumed or provided power at the nodes depends on the connected units/loads, i.e.,
\begin{align}\label{eq:units_to_bus_constraint}
    p_{g}(k) = U [p_t(k)^T~p_s(k)^T~p_r(k)^T~p_d(k)]^T
\end{align}
where $U \in \mathbb{B}^{5 \times 7}$ is a matrix with entries
\begin{equation*}
    U_{ij} = \begin{cases}
1, & \text{if unit/load } j \text{ is connected to node } i, \\
0, & \text{otherwise}. 
\end{cases}
\end{equation*}

\subsection{Cost}

The cost $ \ell \in \mathbb{R}$ to operate the \ac{mg} at time instant $k$ is
\begin{align}\label{eq:stage_cost_sum}
    \ell(k) = \ell_t^{sw}(k) + \ell_p(k) + \ell_x(k) + \ell_{loss}(k).
\end{align}
It is composed of the following parts.

With $c_0,c_1\in \mathbb{R}_{\geq 0}^{2}$,
the costs of enabling/disabling conventional units plus power-independent running costs are
\begin{align}
    \ell_t^{sw}(k) = c_0^T |\delta_{t}(k) - \delta_{t}(k-1)| + c_1^T \delta_{t}(k),
\end{align} 

The cost $\ell_p$ models fuel costs, incentivizes \ac{res} infeed and accounts for storage conversion losses, i.e.,
\begin{align}
    \ell_p(k) =  c_2^T  p_{t}(k) + c_3^T p_{r}(k) + c_4^T |p_{s}(k)|,
\end{align}
with $c_2 \in \mathbb{R}_{\geq 0}^{2}$, $c_3 \in \mathbb{R}_{\leq 0}^{2}$ and $c_4 \in \mathbb{R}_{\geq 0}^{2}$.

To penalize deviations of the stored energy $x(k)$ outside the permissible range $[\underline{x}, \overline{x}]$, we use $c_5\in\mathbb{R}_{\geq 0}^{2}$ and formulate
\begin{align}
    \ell_x(k) = c_5^T \left( \max(0, \underline{x} - x(k)) + \max(0, x(k) - \overline{x}) \right),
\end{align}

Finally, the cost $\ell_{loss}$ is modeled with $c_6\in\mathbb{R}_{\geq 0}$ as
\begin{align}\label{eq:constraint_losses}
    \ell_{loss}(k) = c_6 p_{loss}(k).
\end{align}

\subsection{Model Predictive Control Problems}

The decision variables of the \ac{mpc} problems are 
\[P= \big[[p_u(k+h|k)^T~p_g(k+h|k)^T~p_e(k+h|k)^T]^T\big]_{h=0}^{H-1}\]
where
$p_u(k)=[p_t(k)^T~p_s(k)^T~p_r(k)^T]^T$, and
$H-1$ is the prediction horizon.
We further define
$\Delta_t=[ \delta_t(k+h|k)]_{\scriptstyle h=0}^{\scriptstyle H-1}$,
$X=[x(k+h|k)]_{h=0}^{H-1}$, and
$\Theta = [\theta(k+h|k)]_{h=0}^{H-1} $.
Finally, discount factor $\gamma\in(0,1)$ is used to prioritize near-future decisions and formulate the following \ac{mpc} problem.

\begin{problem}[Nonlinear \ac{mpc} for \ac{mg} operation]\label{prob:MINLP}
\begin{align*}
    \min_{\substack{{\Theta},P, {\Delta}_t, X }} 
    &\textstyle\sum_{h=0}^{H-1} \ell ( k+h|k ) \gamma^{h}
\end{align*}
subject to \eqref{eq:p_t_ineq}--\eqref{eq:units_to_bus_constraint} as well as
\begin{subequations}
\begin{align}
    p_{g,i}(k+h|k) &= \textstyle\sum_{j\in \mathbb{V}_i} p_{e,ij}(k+h|k) ,\quad \forall i\in\mathbb{V}, \label{eq:data-driven_powerflow_constraint_grid}\\
    p_{e,ij}(k+h|k) &=  \breve{g}_{ij} - \big(\tilde{g}_{ij} \cos(\theta_{ij}(k+h|k)) \nonumber \\
        & \hspace{-1.9em} + \tilde{b}_{ij} \sin(\theta_{ij}(k+h|k)) \big), \quad \forall \{i,j\}\in\mathbb{E},\\
    p_{e,ji}(k+h|k) &=  \breve{g}_{ji} - \big(\tilde{g}_{ij} \cos(-\theta_{ij}(k+h|k)) \nonumber \\
        & \hspace{-1.9em} + \tilde{b}_{ij} \sin(-\theta_{ij}(k+h|k)) \big), \quad \forall \{i,j\}\in\mathbb{E},
\end{align}
for $h=0,\dots,H-1$ with initially available $x(k|k)=x_k$ and $\delta_t(k-1|k) = \delta_{t,k-1}$.
\end{subequations}
\end{problem}

Problem~\ref{prob:MINLP} is a nonconvex, mixed-integer optimization problem with an explicit model of the power lines assuming known parameters. 
In what follows, we employ the data-driven representations from Section~\ref{sec:data-driven-NL-PF}.
Therefore, we use ${ \Phi = [\phi(k+h|k)]_{h=0}^{H-1} }$ with $\phi(k)$ from \eqref{eq:data-driven-aux} and ${\upalpha=[\alpha(k+h|k)]_{h=0}^{H-1}}$, $\alpha(k) \in \mathbb{R}^9$ in the following problem.

\begin{problem}[Nonconvex data-driven \acs{mpc}]\label{prob:dd_mpc_acopf}
\[
    \min_{\substack{\upalpha,\Phi, P, \Delta_t, X}} \textstyle\sum_{h=0}^{H-1} \ell (k+h|k) \gamma^{h}
\]
subject to \eqref{eq:p_t_ineq}--\eqref{eq:units_to_bus_constraint} and \eqref{eq:data-driven_powerflow_constraint_grid} as well as
\begin{subequations}
\begin{equation}\label{eq:data-driven_powerflow_constraint_mpc}
\begin{bmatrix}
    H_1(\{\phi^m(k)\}_{k=n}^{n+N-1}) \\
    H_1(\{p_e^m(k)\}_{k=n}^{n+N-1})
\end{bmatrix} \alpha(k+h|k) = 
\begin{bmatrix}
    {\phi}(k+h|k) \\
    {p}_e(k+h|k)
\end{bmatrix}\hspace{-0.35em},
\end{equation}
\begin{equation}\label{eq:nonconvex_constraint_mpc}
    \phi_i^2(k+h|k) + \phi_{i+1}^2(k+h|k) = 1, \quad \forall i \in [2,4,6,8],
\end{equation}
for $h=0,\dots,H-1$ with $x(k|k)=x_k$, $\delta_t(k-1|k) = \delta_{t,k-1}$.
\end{subequations}
\end{problem}

We further define an \ac{mpc} with a convex data-driven active power flow representation using the ideas from Section~\ref{sec:convex_opf}.
Note that the following convex mixed-integer optimization problem considers active power flow using only \ac{io} data.

\begin{problem}[Convex data-driven \ac{mpc}]\label{prob:convex_dd_mpc_acopf}
\begin{align*}
    \min_{\substack{\upalpha,\Phi, P, \Delta_t, X}} \textstyle\sum_{h=0}^{H-1} \big( \ell (k+h|k) \gamma^{h} -\beta \textstyle\sum\limits_{\hspace{-0.3em} i \in \{2,4,6,8\} \hspace{-0.3em} } \phi_i(k+h|k)\big)
\end{align*}
subject to \eqref{eq:p_t_ineq}--\eqref{eq:units_to_bus_constraint}, \eqref{eq:data-driven_powerflow_constraint_grid} and \eqref{eq:data-driven_powerflow_constraint_mpc} as well as
\begin{equation}\label{eq:convex_constraint_mpc}
    \phi_i^2(k+h|k) + \phi_{i+1}^2(k+h|k) \leq 1, \quad \forall i \in [2,4,6,8],
\end{equation}
for $h=0,\dots,H-1$ with $x(k|k)=x_k$, $\delta_t(k-1|k) = \delta_{t,k-1}$.
\end{problem}


\section{Case Study}
\label{sec:casestudy}

We will now evaluate the closed-loop performance of different \ac{mpc} approaches in a numerical case study.
First, we describe the simulation setup and the parameters. 
Then, we assess the performance over a seven day simulation.

\subsection{Simulation Setup}

\begin{table}[b]
\centering
\caption{Parameters of the \ac{mg} model~\cite{Hans2021}.}
\label{table:microgrid_parameter}
\renewcommand{\arraystretch}{1.1}
\setlength{\tabcolsep}{4pt}
\begin{tabular}{ll l ll l ll}
\toprule
Param. & Value & & Param. & Value & & Param. & Value \\
\cmidrule{1-2} \cmidrule{4-5} \cmidrule{7-8}
$c_0$ & $\begin{bmatrix} 0.2 \\ 0.1 \end{bmatrix}$ &  &
$c_1$ & $\begin{bmatrix} 0.13 \\ 0.07 \end{bmatrix}$ &  &
$c_2$ & $\begin{bmatrix} 1.56 \\ 1.43 \end{bmatrix}$ \\
$c_3$ & $\begin{bmatrix} -0.8 \\ -1.0 \end{bmatrix}$ &  &
$c_4$ & $\begin{bmatrix} 0.1 \\ 0.05 \end{bmatrix}$ &  &
$c_5$ & $\begin{bmatrix} 10^3 \\ 10^3 \end{bmatrix}$ \\
$c_6$ & $1$ & &
$\gamma$ & $0.9$ & &
$g_{ij}$ & $2$ pu \\
$b_{ij}$ & $-20$ pu & &
$p_t^{\text{min}}$ & $\begin{bmatrix} 0.3 \\ 0.1 \end{bmatrix}$ pu &  &
$p_t^{\text{max}}$ & $\begin{bmatrix} 0.9 \\ 0.6 \end{bmatrix}$ pu \\
$p_e^{\text{min}}$ & $[-1_{N_e}]$ pu & &
$p_s^{\text{min}}$ & $\begin{bmatrix} -1 \\ -1 \end{bmatrix}$ pu &  &
$p_s^{\text{max}}$ & $\begin{bmatrix} 1 \\ 1 \end{bmatrix}$ pu \\
$p_e^{\text{max}}$ & $[1_{N_e}]$ pu & &
$A_s$ & $\begin{bmatrix} 1 & 0 \\ 0 & 1 \end{bmatrix}$ & &
$B_s$ & $\begin{bmatrix} \frac{1}{2} & 0 \\ 0 & \frac{1}{2} \end{bmatrix}$ \\
$x^{\text{min}}$ & $\begin{bmatrix} 0 \\ 0 \end{bmatrix}$ pu h & &
$x^{\text{max}}$ & $\begin{bmatrix} 7 \\ 4 \end{bmatrix}$ pu h & &
$x_0$ & $\begin{bmatrix} 0.5 \\ 0.5 \end{bmatrix}$ pu h \\
$\underline{x}$ & $\begin{bmatrix} 0.5 \\ 0.5 \end{bmatrix}$ pu h & &
$\overline{x}$ & $\begin{bmatrix} 6.5 \\ 3.5 \end{bmatrix}$ pu h & &
$\delta_{0,-1}$ & $\begin{bmatrix} 1 \\ 0 \end{bmatrix}$ \\
$H$ & $6$ & &
$\beta$ & $1$ & & & \\
\bottomrule
\end{tabular}
\end{table}

We implemented the case study in Python.
The \ac{mpc} problems were formulated using CVXPY~\cite{diamond2016cvxpy} for convex and Gurobi for nonconvex problems.
All problems were solved using Gurobi~12, which employs different techniques to find the global optima~\cite{gurobi-online}.
Hankel matrices were constructed from previously obtained simulation data.
All computations were executed on an Intel Core i7-8565U CPU with \qty{16}{GB} RAM.
Simulations span seven days at $T_s=\qty{0.5}{h}$ ($K=336$ steps).
An ideal forecaster with perfect knowledge about the future was used to focus on effects associated with the different active power flow representations.
All parameters are given in Table~\ref{table:microgrid_parameter}.
Here, power line data was taken from~\cite{Wolf2024} while all remaining parameters were taken from~\cite{Hans2021}.

\begin{remark}
    Since $c_{3,2} + c_6 = 0$ for the second \ac{res}, the cost function is nondecreasing but not strictly increasing, so Condition~(C1) of Remark~\ref{rem:convex_relaxation} does not hold.
    We therefore set $\beta = 1$.
    Since $\theta_{\max} \leq \qty{0.05}{rad}$, the cost distortion bound evaluates to $\beta N_e \theta_{\max}^2/2 \leq 0.005$, which is small compared to the average optimal cost.
\end{remark}

\subsection{Results}

\begin{figure}

\tikzset{external/export next=true} 

\definecolor{vgMidnightBlue}{RGB}{15, 15, 80}   
\definecolor{vgDarkGreen}{RGB}{0, 80, 0}        
\definecolor{vgDarkRed}{RGB}{120, 0, 0}         
\definecolor{vgDarkOrange}{RGB}{230, 120, 0}    

\definecolor{NewOrangeRed}{RGB}{255, 69, 0}   
\definecolor{DarkNewOrangeRed}{RGB}{178, 34, 0} 

\definecolor{NewPurple}{RGB}{147, 112, 219}  
\definecolor{DarkNewPurple}{RGB}{98, 75, 164} 

\colorlet{colorThermal}{vgDarkBlue}
\colorlet{colorStorage}{vgOrange}
\colorlet{colorRes}{vgGreen}
\colorlet{colorDemand}{vgLightBlue}
\colorlet{colorGrid}{vgRed}

\colorlet{colorThermal1}{vgDarkBlue}
\colorlet{colorThermal2}{vgMidnightBlue}
\colorlet{colorStorage1}{vgOrange}
\colorlet{colorStorage2}{vgRed}
\colorlet{colorRes1}{vgGreen}
\colorlet{colorRes2}{vgDarkGreen}
\colorlet{colorLine1}{vgRed}
\colorlet{colorLine2}{vgDarkRed}

\pgfplotsset{linestyleThermal1 stacked/.style = {%
  color=colorThermal1,
  fill=colorThermal1!20!white,
  on layer=axis foreground,
  line width=0.65pt,
}}

\pgfplotsset{linestyleThermal2 stacked/.style = {%
  color=colorThermal2,
  fill=colorThermal2!20!white,
  on layer=axis foreground,
  line width=0.65pt,
}}

\pgfplotsset{linestyleStoragePower1 stacked/.style = {%
  color=colorStorage1,
  fill=colorStorage1!20!white,
  on layer=axis foreground,
  line width=0.65pt,
}}

\pgfplotsset{linestyleStoragePower2 stacked/.style = {%
  color=colorStorage2,
  fill=colorStorage2!20!white,
  on layer=axis foreground,
  line width=0.65pt,
}}

\pgfplotsset{linestyleGridPower stacked/.style = {%
  color=colorGrid,
  fill=colorGrid!20!white,
  on layer=axis foreground,
  line width=0.65pt,
}}

\pgfplotsset{linestyleRes1 stacked/.style = {%
  color=colorRes1,
  fill=colorRes1!20!white,
  on layer=axis foreground,
  line width=0.65pt,
}}

\pgfplotsset{linestyleRes2 stacked/.style = {%
  color=colorRes2,
  fill=colorRes2!20!white,
  on layer=axis foreground,
  line width=0.65pt,
}}

\pgfplotsset{linestyleDemand stacked/.style = {%
  color=colorDemand,
  fill=colorDemand!20!white,
  on layer=axis foreground,
  line width=0.65pt,
}}

\pgfplotsset{limit/.style = {const plot, color=black, dash dot, thin}}

\pgfplotsset{
    /pgfplots/layers/Bowpark/.define layer set={
        axis background,axis grid,main,axis ticks,axis lines,axis tick labels,
        axis descriptions,axis foreground
    }{/pgfplots/layers/standard},
}

\pgfplotsset{resultsPlot/.style={%
    clip = false,
    minor x tick num=3,
    set layers=Bowpark,
    grid=both,
    grid style={draw=black!25},
    major tick length=0pt,
    minor tick length=0pt,
    axis lines = left,
    axis line style= {-, draw opacity=0.0},
    y tick label style={
        /pgf/number format/.cd,
            scaled y ticks = false,
            fixed,
            precision=3,
        /tikz/.cd
        },
    height = 38mm,
    xtick = {0, 48, ..., 336},
    xticklabels = {0, 1, ..., 7},
    xmin = 0,
    xmax = 336,
		clip=false,
    width=90mm,
    legend columns=4,
    legend style={
      at={(0.55, 1.1)},
      anchor=south,
      draw=none,
      fill=none,
      legend cell align=left,
      text width=10mm,
      /tikz/every even column/.append style={column sep=1mm}
      },
	}
}

\pgfplotsset{resultsPlotPower/.style={%
    resultsPlot,
    stack plots=y, stack negative=separate, area style, enlarge x limits=false,
    xlabel = {},
    ymin = -2.5,
    ymax = 2.5,
    ytick={-4, -2, 0, 2, 4},
    minor y tick num=3,
  },
}

\begin{tikzpicture}[font=\footnotesize, scale=0.95]

\draw[draw=none, fill=none] (-1.02, 2.3) rectangle (7.44, -6.85);

\begin{axis}[%
  resultsPlotPower,
  xticklabels = {~},
  ylabel={Unit power in $\unit{pu}$},
  ]

  \addplot [linestyleThermal1 stacked] table [x expr=\coordindex, y=Conv. 1 Power, col sep=comma] {data/eval_dd_convex_acopf_nonlinearMG.csv} \closedcycle;
  \addlegendentry{Conv. 1}

      \addplot [linestyleThermal2 stacked] 
        table [x expr=\coordindex, y=Conv. 2 Power, col sep=comma] 
        {data/eval_dd_convex_acopf_nonlinearMG.csv} \closedcycle;
    \addlegendentry{Conv. 2}

    \addplot [linestyleStoragePower1 stacked] table [x expr=\coordindex, y=Storage 1 Power, col sep=comma] {data/eval_dd_convex_acopf_nonlinearMG.csv} \closedcycle;
  \addlegendentry{BESS 1}

  \addplot [linestyleStoragePower2 stacked] 
    table [x expr=\coordindex, y=Storage 2 Power, col sep=comma] 
    {data/eval_dd_convex_acopf_nonlinearMG.csv} \closedcycle;
    \addlegendentry{BESS 2}

  \addplot [linestyleRes1 stacked] table [x expr=\coordindex, y=Wind Power, col sep=comma] {data/eval_dd_convex_acopf_nonlinearMG.csv} \closedcycle;
  \addlegendentry{RES 1}

  \addplot [linestyleRes2 stacked] 
    table [x expr=\coordindex, y=PV Power, col sep=comma] 
    {data/eval_dd_convex_acopf_nonlinearMG.csv} \closedcycle;
  \addlegendentry{RES 2}

  \addplot [linestyleDemand stacked] table [x expr=\coordindex, y=Load, col sep=comma] {data/eval_dd_convex_acopf_nonlinearMG.csv} \closedcycle;
  \addlegendentry{Load}

\end{axis}

\begin{axis}[%
  shift={(0mm, -26mm)},
  resultsPlot,
  ylabel = {Energy in $\unit{pu\,h}$},
  xticklabels = {~},
  ymin = 0.5,
  ymax = 3.5,
  ytick = {0.5, 3.5},
  minor y tick num=2,
  ]

    \addplot [color = vgOrange, line width=0.65pt] table [x expr=\coordindex, y=Stored Energy 1, col sep=comma] {data/eval_dd_convex_acopf_nonlinearMG.csv};

    \addplot [color = vgRed, line width=0.65pt] table [x expr=\coordindex, y=Stored Energy 2, col sep=comma] {data/eval_dd_convex_acopf_nonlinearMG.csv};

\end{axis}

\begin{axis}[%
  shift={(0mm, -60mm)},
  resultsPlot,
  ylabel = {Line power in $\unit{pu}$},
  ymin = -1,
  ymax = 1,
  ytick = {-1, 0, 1},
  minor y tick num=3,
  xlabel = {Time in days},
  ]

    \addplot [color = NewPurple, line width=0.65pt] table [x expr=\coordindex, y=Line Power 24, col sep=comma] {data/eval_dd_convex_acopf_nonlinearMG.csv};
    \addlegendentry{$p_{e,24}$}

    \addplot [color = DarkNewPurple, line width=0.65pt] table [x expr=\coordindex, y=Line Power 42, col sep=comma] {data/eval_dd_convex_acopf_nonlinearMG.csv};
    \addlegendentry{$p_{e,42}$}

     \addplot [color = vgLightBlue, line width=0.65pt] table [x expr=\coordindex, y=Line Power 25, col sep=comma] {data/eval_dd_convex_acopf_nonlinearMG.csv};
    \addlegendentry{$p_{e,25}$}

     \addplot [color = vgDarkBlue, line width=0.65pt] table [x expr=\coordindex, y=Line Power 52, col sep=comma] {data/eval_dd_convex_acopf_nonlinearMG.csv};
    \addlegendentry{$p_{e,52}$}

     \addplot [color = vgGreen, line width=0.65pt] table [x expr=\coordindex, y=Line Power 35, col sep=comma] {data/eval_dd_convex_acopf_nonlinearMG.csv};
    \addlegendentry{$p_{e,35}$}

     \addplot [color = vgDarkGreen, line width=0.65pt] table [x expr=\coordindex, y=Line Power 53, col sep=comma] {data/eval_dd_convex_acopf_nonlinearMG.csv};
    \addlegendentry{$p_{e,53}$}

    \addplot [color = NewOrangeRed, line width=0.65pt, densely dashed] table [x expr=\coordindex, y=Line Power 12, col sep=comma] {data/eval_dd_convex_acopf_nonlinearMG.csv};
    \addlegendentry{$p_{e,12}$}

    \addplot [color = DarkNewOrangeRed, line width=0.65pt, densely dotted] table [x expr=\coordindex, y=Line Power 21, col sep=comma] {data/eval_dd_convex_acopf_nonlinearMG.csv};
    \addlegendentry{$p_{e,21}$}

\end{axis}

\end{tikzpicture}
    \vspace{-1.5em}
    \caption{Simulation results of controlling the reference \ac{mg} using the convex data-driven \ac{mpc} over seven days. The unit power is presented as a stacked area plot.}
    \label{fig:simulation}
\end{figure}

\begin{figure}

\tikzset{external/export next=true}

\begin{tikzpicture}[font=\footnotesize]
  \begin{semilogxaxis}[ 
    myPlot,
    clip=false,
    height = 28mm,
    width = 78mm,
    xmin = 0.01,
    xmax = 10,
    line width=0.7pt,
    xtick = {1e-2, 1e-1, 1e0, 1e1},
    xticklabels = {0.01, 0.1, 1, 10},
    y axis line style={white},
    ytick style={draw=none},
    yticklabels={},
    ymax = 3.5,
    xlabel = {Solve time in s},
    ]

\addplot [linestyle boxplot=purple] 
  table[y = Solve Time, col sep=comma]{data/eval_acopf_nonlinearMG_solveTime.csv};

\addplot [linestyle boxplot=vgDarkBlue] 
  table[y = Solve Time, col sep=comma]{data/eval_dd_nonconvex_acopf_nonlinearMG_solveTimes.csv};

\addplot [linestyle boxplot=vgLightBlue] 
  table[y = Solve Time, col sep=comma]{data/eval_dd_convex_acopf_nonlinearMG_solveTimes.csv};

  \node[anchor=east] at (axis cs:0.01, 1) {Model-based};
  \node[anchor=east] at (axis cs:0.01, 2) {Nonconvex DD};
  \node[anchor=east] at (axis cs:0.01, 3) {Convex DD};

  \end{semilogxaxis}
\end{tikzpicture}
    \vspace{-1.5em}
    \caption{Solve times across all $336$ time steps for the model-based Problem~\ref{prob:MINLP}, the nonconvex data-driven Problem~\ref{prob:dd_mpc_acopf}, and the convex data-driven Problem~\ref{prob:convex_dd_mpc_acopf}.
    The large dots are the median, the surrounding white area the interquartile range, and the whiskers extend to values within $1.5$ times the interquartile range.
    Values beyond the whiskers are outliers.}
    \label{fig:boxplot}
\end{figure}

Figure \ref{fig:simulation} shows the results of simulations with Problem~\ref{prob:dd_mpc_acopf}. 
Simulations for Problems~\ref{prob:MINLP} and~\ref{prob:convex_dd_mpc_acopf} were also performed.
For all of them, the \ac{mpc} problems led to identical output trajectories and identical costs (within errors bounds of $10^{-4}$) of $\bar{\ell}_o = \textstyle\frac{1}{K} \textstyle\sum_{k=1}^{K} \left(\ell_t^{sw} (k)  + \ell_p (k) \right)=-0.2602$ as well as $\bar{\ell}_{loss} = \textstyle\frac{1}{K} \textstyle\sum_{k=1}^{K} \ell_{loss} (k)=0.0047$. 
The identical optimal results (within the specified numerical precision) indicate that the data-driven formulations can successfully replace the physics-based ones.
The identical cost across all formulations further indicates that the regularization term with $\beta = 1$ introduces negligible cost distortion (see also Remark~\ref{rem:convex_relaxation}).

Figure~\ref{fig:boxplot} shows the distribution of the solve times for all $336$ time steps.
The convex data-driven \ac{mpc} achieves the shortest median solve time of $0.08\unit{s}$, followed by the reference \ac{mpc} at $0.18\unit{s}$ and the nonconvex data-driven \ac{mpc} at $0.33\unit{s}$, which has the highest median solve time.
Given a sampling time of $30$ minutes, the maximum solve times are well within the real-time requirements of the system at hand, demonstrating the applicability of the proposed approaches.


\section{Conclusions}\label{sec:conclusion}

We proposed data-driven formulations of nonlinear active power flows based on behavioral systems theory. 
Our approaches enable active power flow optimization without relying on a physics-based model and require no explicit system identification step.
Simulations demonstrated that all approaches yield identical output trajectories up to a numerical precision of $10^{-4}$, while the required time to solve the optimization problem varies.

Future research will evaluate scalability and focus on extending the proposed method to meshed networks as well as topology-agnostic formulations.
Moreover, convex relaxations will be further investigated.

\bibliographystyle{IEEEtran}
\bibliography{manual}

\begin{thebibliography}{10}
\providecommand{\url}[1]{#1}
\csname url@samestyle\endcsname
\providecommand{\newblock}{\relax}
\providecommand{\bibinfo}[2]{#2}
\providecommand{\BIBentrySTDinterwordspacing}{\spaceskip=0pt\relax}
\providecommand{\BIBentryALTinterwordstretchfactor}{4}
\providecommand{\BIBentryALTinterwordspacing}{\spaceskip=\fontdimen2\font plus
\BIBentryALTinterwordstretchfactor\fontdimen3\font minus \fontdimen4\font\relax}
\providecommand{\BIBforeignlanguage}[2]{{%
\expandafter\ifx\csname l@#1\endcsname\relax
\typeout{** WARNING: IEEEtran.bst: No hyphenation pattern has been}%
\typeout{** loaded for the language `#1'. Using the pattern for}%
\typeout{** the default language instead.}%
\else
\language=\csname l@#1\endcsname
\fi
#2}}
\providecommand{\BIBdecl}{\relax}
\BIBdecl

\bibitem{Bordons2020}
C.~Bordons, \emph{Model predictive control of microgrids}, ser. Advances in industrial control, F.~Garcia-Torres and M.~A. Ridao, Eds.\hskip 1em plus 0.5em minus 0.4em\relax Springer, 2020.

\bibitem{Low2014}
S.~H. Low, ``Convex relaxation of optimal power flow—part i: Formulations and equivalence,'' \emph{IEEE Trans. Control Netw. Syst.}, vol.~1, no.~1, pp. 15--27, 2014.

\bibitem{Yu2025}
N.~Yu, S.~Zhang, J.~Qin, P.~Hidalgo-Gonzalez, R.~Dobbe, Y.~Liu, A.~Dubey, Y.~Wang, J.~Dirkman, H.~Zhong, N.~Lu, E.~Ma, Z.~Ding, D.~Cao, J.~Zhao, and Y.~Gao, ``Data-driven control, optimization, and decision-making in active power distribution networks,'' \emph{Applied Energy}, vol. 397, p. 126253, 2025.

\bibitem{Liu2019}
Y.~Liu, N.~Zhang, Y.~Wang, J.~Yang, and C.~Kang, ``Data-driven power flow linearization: A regression approach,'' \emph{IEEE Trans. Smart Grid}, vol.~10, no.~3, pp. 2569--2580, 2019.

\bibitem{Guo2022}
L.~Guo, Y.~Zhang, X.~Li, Z.~Wang, Y.~Liu, L.~Bai, and C.~Wang, ``Data-driven power flow calculation method: A lifting dimension linear regression approach,'' \emph{IEEE Trans. Power Syst.}, vol.~37, no.~3, pp. 1798--1808, 2022.

\bibitem{Polderman1998}
J.~W. Polderman and J.~C. Willems, \emph{Introduction to Mathematical Systems Theory}.\hskip 1em plus 0.5em minus 0.4em\relax Springer, 1998.

\bibitem{Willems2004}
J.~Willems, I.~Markovsky, P.~Rapisarda, and B.~De~Moor, ``A note on persistency of excitation,'' in \emph{43rd CDC}, 2004.

\bibitem{Berberich2020}
J.~Berberich and F.~Allgöwer, ``A trajectory-based framework for data-driven system analysis and control,'' in \emph{ECC}, 2020, pp. 1365--1370.

\bibitem{Markovsky2023}
I.~Markovsky, L.~Huang, and F.~Dörfler, ``Data-driven control based on the behavioral approach: From theory to applications in power systems,'' \emph{IEEE Control Syst.}, vol.~43, no.~5, pp. 28--68, 2023.

\bibitem{Lipka2024}
J.~B. Lipka and C.~A. Hans, ``Data-driven model predictive control of battery storage units,'' in \emph{ACC}, 2024, pp. 2230--2235.

\bibitem{Bilgic2022}
D.~Bilgic, A.~Koch, G.~Pan, and T.~Faulwasser, ``Toward data-driven predictive control of multi-energy distribution systems,'' \emph{Electric Power Systems Research}, vol. 212, p. 108311, 2022.

\bibitem{Molodchyk2025a}
O.~Molodchyk, P.~Schmitz, A.~Engelmann, K.~Worthmann, and T.~Faulwasser, ``Towards data-driven multi-stage opf,'' in \emph{2025 IEEE Kiel PowerTech}.\hskip 1em plus 0.5em minus 0.4em\relax IEEE, 2025, pp. 1--6.

\bibitem{Coulson2019}
J.~Coulson, J.~Lygeros, and F.~Dörfler, ``Data-enabled predictive control: In the shallows of the deepc,'' in \emph{ECC}, 2019.

\bibitem{Diestel2017}
R.~Diestel, \emph{Graph Theory}, fifth edition~ed., ser. Graduate texts in mathematics.\hskip 1em plus 0.5em minus 0.4em\relax Berlin: Springer, 2017, no. 173.

\bibitem{Kundur2007}
P.~Kundur, \emph{Power system stability and control}.\hskip 1em plus 0.5em minus 0.4em\relax McGraw-Hill, 2007.

\bibitem{dorfler_kron_2013}
F.~Dörfler and F.~Bullo, ``Kron {Reduction} of {Graphs} {With} {Applications} to {Electrical} {Networks},'' \emph{IEEE Trans. Circuits Syst. I Regul. Pap.}, vol.~60, no.~1, pp. 150--163, 2013.

\bibitem{Lasseter}
R.~Lasseter, ``Microgrids,'' in \emph{IEEE PES WM}, 2002, pp. 305--308.

\bibitem{Jabr2006}
R.~Jabr, ``Radial distribution load flow using conic programming,'' \emph{IEEE Trans. Power Syst.}, vol.~21, no.~3, pp. 1458--1459, 2006.

\bibitem{Low2014a}
S.~H. Low, ``Convex relaxation of optimal power flow---part {II}: Exactness,'' \emph{IEEE Trans. Control Netw. Syst.}, vol.~1, no.~2, pp. 177--189, Jun. 2014.

\bibitem{Hans2021}
C.~A. Hans, \emph{Operation control of islanded microgrids}.\hskip 1em plus 0.5em minus 0.4em\relax Shaker Verlag, 2021.

\bibitem{diamond2016cvxpy}
S.~Diamond and S.~Boyd, ``{CVXPY}: {A} {P}ython-embedded modeling language for convex optimization,'' \emph{J. Mach. Learn. Res.}, vol.~17, no.~83, pp. 1--5, 2016.

\bibitem{gurobi-online}
\BIBentryALTinterwordspacing
{Gurobi Optimization, Inc.} (2026) Gurobi optimizer: Nonlinear features. [Online]. Available: \url{https://docs.gurobi.com/projects/optimizer/en/current/features/nonlinear.html}
\BIBentrySTDinterwordspacing

\bibitem{Wolf2024}
H.~M.~H. Wolf and C.~A. Hans, ``Identification of power systems with droop-controlled units using neural ordinary differential equations,'' in \emph{ECC}, 2025, pp. 800--806.

\end{thebibliography}

\end{document}